\documentclass[aps,showpacs,floatfix,preprintnumbers]{revtex4}

\usepackage{amssymb}
\usepackage{epstopdf}
\usepackage{graphicx}
\usepackage{dcolumn}
\usepackage{bm}
\usepackage{amsmath}
\usepackage[usenames]{color}
\usepackage{epstopdf}

\usepackage{graphicx, epsfig}
\usepackage{color}
\usepackage{amsmath}
\usepackage{dcolumn}
\newcommand{\be}{\begin{equation}}
\newcommand{\ee}{\end{equation}}
\newcommand{\bea}{\begin{eqnarray}}
\newcommand{\eea}{\end{eqnarray}}

\newcommand{\gapp}{\mathrel{\raise.3ex\hbox{$>$}\mkern-14mu \lower0.6ex\hbox{$\sim$}}}
\newcommand{\lapp}{\mathrel{\raise.3ex\hbox{$<$}\mkern-14mu \lower0.6ex\hbox{$\sim$}}}
\def\bbox{{\,\lower0.9pt\vbox{\hrule \hbox{\vrule height 0.2 cm
\hskip 0.2 cm \vrule  height 0.2 cm}\hrule}\,}}

\begin{document}
\title{Extending cascading gravity model to lower dimensions}
\author{Peng Hao$^1$}
\author{Dejan Stojkovic$^{1,2}$}
\affiliation{$^1$ HEPCOS, Department of Physics, SUNY at Buffalo, Buffalo, NY 14260-1500}
\affiliation{$^2$ Perimeter Institute for Theoretical Physics, 31 Caroline St. N., Waterloo, ON, N2L 2Y5, Canada}
\date{\today}
\begin{abstract}
The cascading gravity model was proposed to eliminate instabilities of the original DGP model by embedding our 4D universe into a $5D$ brane, which is itself embedded in a $6D$ bulk. Thus gravity cascades from $6D$ down to $4D$ as we decrease the length scales.  We show that it is possible to extend this setup to lower dimensions as well, i.e. there is a self-consistent embedding of a $3D$ brane into a $4D$ brane, which is itself embedded in a $5D$ bulk and so on. This extension fits well into the ``vanishing dimensions" framework in which dimensions open up as we increase the length scales.
\end{abstract}
\maketitle

\section{Introduction}
 It has been argued for quite some time that our theory of gravity embodied in Einstein's general relativity needs to be modified on large cosmological scales where we run into problems like the cosmological constant problem, dark matter problem etc. One of the most interesting non-trivial modifications is the so-called DGP model \cite{a}. In that model, our universe is just a $4D$ dimensional brane embedded in a 5D space-time (where $D$ refers to the number of space-time dimensions). Unlike the original model of large extra dimensions, where extra dimensions are either compacted or warped, the bulk space of DGP model is infinite.  Fine tuning between the bulk cosmological constant and brane tension ensures that gravity appears $4D$ on reasonable scales, and becomes $5D$ only on very large cosmological scales. Such a setup was extensively studied in context of modified gravity and cosmology.

 As a very interesting extension of the DGP model, the so-called cascading gravity model was proposed in \cite{d,g,h,Kaloper:2007ap}. The main motivation for it was the apparent instability of the original DGP model. In the cascading model, the appearance of ghosts is prevented by embedding the $4D$ brane in a higher dimensional $5D$ brane, which is itself embedded in a $6D$ bulk.
 Thus, new dimensions open up for gravity as we increase the length scale. A priori, one does not have to stop with a $6D$ bulk, and can extend this cascading setup to even higher dimensions.

On the other hand, the cascading gravity model could be extended from the left, i.e. from the lower dimensional side. Apart from the pure academic interest, one  motivation for this comes from the so-called ``vanishing" or ``evolving" dimensions models \cite{Anchordoqui:2010er,Anchordoqui:2010hi,Stojkovic:2013lga,review,Mureika:2011bv}. In ``vanishing dimensions", the short distance physics is lower dimensional rather than higher dimensional. Having less dimensions at high energies has manifold advantages, as pointed out in \cite{Anchordoqui:2010er}. Reducing the number of dimensions at short distances can be achieved by an ad hoc ordered lattice model \cite{Anchordoqui:2010er,Anchordoqui:2010hi}, or elaborate stringy model \cite{as}. It would be interesting to see if the cascading gravity model can also be extended to lower dimensions, which is the main goal of this paper.

\section{Brief review of the $DGP$ model}

To setup the formalism, in this section we present a brief pedagogical description of the relevant details of the DGP model, filling the gaps in calculations that were left in the original paper. By doing so, we will facilitate the presentation of analog calculations in lower dimensions after Section III.

\subsection{Linearized Einstein equations in $4D$ and $5D$}
In the following sections we will use linearized Einstein equations in $4D$ and $5D$. For convenience we present them here.
We begin with  Einstein equations in $4D$ space-time.

\begin{equation}
\label{Ein}
	G_{\mu\nu}=8\pi GT_{\mu\nu}
\end{equation}

If we perturb the metric around the flat space-time, i.e. $g_{\mu\nu}=\eta_{\mu\nu}+h_{\mu\nu}$, and keep only the linear terms,
the right hand side becomes
\begin{equation}
\label{eqn1.2}
\begin{split}
	G_{\mu\nu}&=R_{\mu\nu}-\frac{1}{2}\eta_{\mu\nu}R  \\
          &=\frac{1}{2}(\partial_\sigma\partial_\nu h^\sigma{}_\mu + \partial_\sigma\partial_\mu h^\sigma{}_\nu -\partial_\mu \partial_\nu h-\square h_{\mu\nu}-\eta_{\mu\nu}\partial_\rho\partial_\lambda h^{\rho\lambda}+\eta_{\mu\nu} \square h) ,
\end{split}
\end{equation}
where $h$ is trace of $h_{\mu\nu}$ in 4D, i.e. $h \equiv h_{\mu}^{\mu}$.
We can simplify this expression in the harmonic gauge
\begin{equation}
\begin{split}
	\partial_\mu h^{\mu\nu}=\frac{1}{2}\partial^\nu h
\end{split}
\end{equation}
The first term in \eqref{eqn1.2} becomes
\begin{equation}
\begin{split}
	\partial_\sigma \partial_\nu h^\sigma{}_\mu &=\eta_{\mu\nu} \partial_\nu \partial_\sigma h^{\sigma\lambda} \\
	&=\frac{1}{2} \partial_\mu \partial_\nu h
\end{split}
\end{equation}
Similarly,
\begin{equation}
\partial_\sigma \partial_\mu h^\sigma{}\nu=\frac{1}{2}\partial_\mu \partial_\nu h
\end{equation}
Thus, the first three terms in \eqref{eqn1.2} add up to zero.
Furthermore, \[\eta_{\mu\nu}\partial_\rho\partial_\lambda h^{\rho\lambda}=\eta_{\mu\nu}\partial_\rho \frac{1}{2}\partial^\rho h=\frac{1}{2}\eta_{\mu\nu}\square h\]
When we sum all terms in \eqref{eqn1.2} we get

\begin{equation}
\square h_{\mu\nu}-\frac{1}{2}\eta_{\mu\nu}\square h=-16\pi GT_{\mu\nu}\end{equation}

In $5D$, the linearized Einstein tensor takes the form,

\begin{equation}
\label{5deinsteintersor}
G_{AB}=-\frac{1}{2}(\square_5 h_{AB}-\frac{1}{2}\eta_{AB}\square_5 h^C_C)
\end{equation}
where $h^C_C$ is trace of $h_{AB}$ in 5D.
Greek letters $\mu$, $\nu$, $\lambda$ $\cdots$ denote $4D$ space-time coordinates, while Latin letters  $A$, $B$, $C$, $\cdots$ go over the whole $5D$ space-time. The letter $y$ is reserved for the extra dimension ($5$th coordinate).

\subsection{The $DGP$ action}
\label{subsection2b}

We can now write down the original $DGP$ action,
\begin{equation} \label{dgpaction}
S= \frac{M_5^3}{2}\int{\mathrm{d}^5x\,\sqrt{-G}R_5}+\frac{M_4^2}{2}\int{\mathrm{d}^4x\,\sqrt{-g}R}+\int{\mathrm{d}^4x\,\mathcal{L}_{matter}}
\end{equation}
where the $G$ is the determinant of $5D$ metric, while the induced metric on the brane is denoted by
\begin{equation}
 g_{\mu\nu}(x)\equiv G_{\mu\nu} (x,y=0)
\end{equation}

The full $5D$ metric tensor has components
\begin{equation}G_{AB}=\left( \begin{array}{cc}
G_{\mu\nu}&G_{\mu5}\\
G_{5\nu}&G_{55}\end{array} \right)
\end{equation}
As usual, we assume that all the matter fields are confined to the brane
\begin{equation} T_{AB}=\left(\begin{array}{cc}
\delta(y)T_{\mu\nu}&0\\
0&0 \end{array} \right)
\end{equation}
If we vary action in Eq.~(\ref{dgpaction}) with respect to the metric $G_{AB}$ we get the equations of motion
\begin{equation}\label{5DEOM}
M_5^3 G_{AB}+\delta(y)M_4^2G_{\mu\nu}\delta_A^\mu\delta_B^\nu =\delta(y)T_{\mu\nu}\delta_A^\mu\delta_B^\nu
\end{equation}

We can now impose the harmonic gauge in the bulk
\begin{equation}
\label{gauge} \partial^A h_{AB}=\frac{1}{2}\partial_B h^C_C  ,
\end{equation}
and  derive the linearized equations for the induced metric in the brane.

We want to check if we can simplify our calculations by setting $h_{\mu5}=0$, since $T_{\mu5}=0$ .
With the use of  Eq.~\eqref{5deinsteintersor}, Eq.~\eqref{5DEOM} becomes
\begin{equation}
-\frac{M_5^3}{2}(\square_5 h_{\mu5}-\frac{1}{2}\eta_{\mu5}\square_5h^C_C)+\delta(y)M_4^2 G_{\alpha\beta}\delta^\alpha_\mu\delta_5^\beta =\delta(y)T_{\alpha\beta}\delta^\alpha_\mu\delta_5^\beta
\end{equation}

\begin{equation}
\square_5 h_{\mu5}=0
\end{equation}

From here we can conclude that $h_{\mu5}=0$ is a self-consistent choice.
The surviving components of $h_{AB}$ are $h_{\mu\nu}$ and $h_{55}$. With $h_{\mu5}=0$, the gauge condition \eqref{gauge} tells us that
the $h_{55}$ obeys
\begin{equation}
\begin{split}
&\partial^5h_{55} = \frac{1}{2}\partial_5(h^\alpha_\alpha+h^5_5)\\
&\partial_5(h^\alpha_\alpha-h^5_5) =0 \\
&h^\alpha_\alpha =h^5_5
\label{55}
\end{split}
\end{equation}
while $h_{\mu\nu}$ obeys
\begin{equation}
\label{munu}
\partial^\mu h_{\mu\nu} =\frac{1}{2}\partial_\nu (h^\alpha_\alpha+h^5_5)\\
\end{equation}

\begin{equation}
\partial^\mu h_{\mu\nu}=\partial_\nu h^\alpha_\alpha
\label{newgauge}
\end{equation}
The above is the gauge condition for the induced metric on the brane. Now we can rewrite the Einstein's equation for the brane. In this gauge, $G_{\mu\nu}$ takes the form
\begin{equation}
G_{\mu\nu}=-\frac{1}{2}(\square h_{\mu\nu} -\partial_\mu\partial_\nu h^\alpha_\alpha)
\end{equation}
and \eqref{5DEOM} becomes,
\begin{equation}
-\frac{M^3_5}{2}\square_5 (h_{AB}-\frac{1}{2}\eta_{AB}h^C_C)-\frac{\delta(y)}{2}M^2_4(\square h_{\mu\nu}-\partial_\mu\partial_\nu h^\alpha_\alpha)\delta^\mu_A\delta^\nu_B =\delta(y) T_{\mu\nu}\delta^\mu_A\delta^\nu_B
\end{equation}
with $h^\alpha_\alpha=h^5_5$
\begin{equation}
\label{24}
-\frac{M^3_5}{2}\square_5 (h_{AB}-\eta_{AB}h^\alpha_\alpha)-\frac{\delta(y)}{2}M^2_4(\square h_{\mu\nu}-\partial_\mu\partial_\nu h^\alpha_\alpha)\delta^\mu_A\delta^\nu_B =\delta(y) T_{\mu\nu}\delta^\mu_A\delta^\nu_B
\end{equation}
Take the trace,
\begin{equation}
-\frac{M^3_5}{2}\square_5 (h^B_B-5h^\alpha_\alpha)-\frac{\delta(y)}{2}M^2_4(\square h^\alpha_\alpha-\partial_\mu\partial^\mu h^\alpha_\alpha) =\delta(y) T^\alpha_\alpha
\end{equation}
\begin{equation}
\label{25}
-\frac{3}{2}M^3_5\square_5 h^\alpha_\alpha = \delta(y) T^\alpha_\alpha
\end{equation}
Let's look at the $(\mu\nu)$ components of \eqref{24},
\begin{equation}
-\frac{M^3_5}{2}\square_5 (h_{\mu\nu}-\eta_{\mu\nu}h^\alpha_\alpha)-\frac{\delta(y)}{2}M^2_4(\square h_{\mu\nu}-\partial_\mu\partial_\nu h^\alpha_\alpha) =\delta(y) T_{\mu\nu}
\end{equation}
with \eqref{25},
\begin{equation}
-\frac{1}{2}(M^3_5\square_5+\delta(y)M^2_4\square)h_{\mu\nu} =\delta(y)T_{\mu\nu}+\eta_{\mu\nu}-\frac{M^3_5}{2} \square h^\alpha_\alpha -\frac{\delta(y)}{2}M^2_4\partial_\mu\partial_\nu h^5_5
\end{equation}

\begin{equation}
\label{23inDGP}
-\frac{1}{2}(M^3_5\square_5+\delta(y)M^2_4\square)h_{\mu\nu} =(T_{\mu\nu}-\frac{1}{3}\eta_{\mu\nu}T^\alpha_\alpha)\delta(y) -\frac{\delta(y)}{2}M^2_4\partial_\mu\partial_\nu h^5_5
\end{equation}

Massive and massless gravitons have different tensor structures \cite{b,c}. In $4D$ space-time, the massless graviton propagator is
\begin{equation}
 \label{massless} \frac{1}{2}\eta^{\mu\alpha}\eta^{\nu\beta}+\frac{1}{2}\eta^{\mu\beta}\eta^{\nu\alpha}-\frac{1}{2}\eta^{\mu\nu}\eta^{\alpha\beta} ,
 \end{equation}
while the massive graviton propagator is
\begin{equation} \label{massive} \frac{1}{2}\eta^{\mu\alpha}\eta^{\nu\beta}+\frac{1}{2}\eta^{\mu\beta}\eta^{\nu\alpha}-\frac{1}{3}\eta^{\mu\nu}\eta^{\alpha\beta}\end{equation}

The tensor structure we obtained for the induced brane metric in \eqref{23inDGP} appears to be a massive graviton propagator. We can rewrite it in a form similar to the massless graviton
\begin{align}
(T_{\mu\nu}-\frac{1}{3}\eta_{\mu\nu}T^\alpha_\alpha)\delta(y) & = (T_{\mu\nu}-\frac{1}{2}\eta_{\mu\nu}T^\alpha_\alpha)\delta(y)+ \frac{1}{6}\eta_{\mu\nu}T^\alpha_\alpha)\delta(y) \\
& = (T_{\mu\nu}-\frac{1}{2}\eta_{\mu\nu}T^\alpha_\alpha)\delta(y)-\frac{1}{2}M^3_5 \eta_{\mu\nu}\square_5 h^\alpha_\alpha
\end{align}
but the second term on the right hand side only reminds us that we are actually dealing with a tensor-scalar theory.

We are now ready to find the exact graviton propagator. In order to do this, we will first use the Fourier-transformed expression
\begin{equation}
h_{\mu\nu}(x,y)\equiv \int{\frac{\mathrm{d^4p}}{(2\pi)^4}e^{ipx}\tilde{h}_{\mu\nu}(P,y)}
\end{equation}
Note that in momentum space, the last term in \eqref{23inDGP} has the form  of $P_\mu P_\nu$, since we are interested in graviton propagation in the lowest order with one tree diagram, when coupled with a conserved energy-momentum tensor $T_{\mu\nu}$, $T'_{\mu\nu}$, we have
\begin{equation}
\label{conservation}
P^\mu T_{\mu\nu}=P^\nu T_{\mu\nu}=P^\alpha T'_{\alpha\beta}=P^\beta T'_{\alpha\beta}=0
\end{equation}
This implies that the last term in \eqref{23inDGP} gives no contribution.

In Euclidean space, \eqref{23inDGP} coupled to a conserved $T'_{\mu\nu}$ reads
\begin{equation}
\label{note34}
(M^3_5\partial_A\partial^A+\delta(y)M^2_4\partial_\mu\partial^\mu)\int{\frac{\mathrm{d^4p}}{(2\pi)^4}e^{ipx}\tilde{h}_{\mu\nu}(P,y)\tilde{T}^{\prime\mu\nu}}=(\int{\frac{\mathrm{d^4p}}{(2\pi)^4}e^{ipx}\tilde{h}_{\mu\nu}\tilde{T}^{\prime\mu\nu}}-\frac{1}{3}\int{\frac{\mathrm{d^4p}}{(2\pi)^4}e^{ipx}\tilde{T}_\alpha^\alpha\eta_{\mu\nu}\tilde{T}^{\prime\mu\nu}})\delta(y)
\end{equation}
The last term in \eqref{23inDGP} vanishes due to \eqref{conservation}.

From \eqref{note34}, we can get
\begin{equation}
[M^3_5(\mathrm{P}^2-\partial_y^2)+M^2_4\delta(y)\mathrm{P}^2]\tilde{h}_{\mu\nu}\tilde{T}^{\prime\mu\nu}=(\tilde{T}_{\mu\nu}\tilde{T}^{\prime\mu\nu}-\frac{1}{3}\tilde{T}^\alpha_\alpha\tilde{T}^{\prime\beta}_\beta)\delta(y)
\end{equation}
where $\mathrm{P}^2=p_4^2+p_1^2+p_2^2+p_3^2$. We can now solve for $\tilde{h}_{\mu\nu}\tilde{T}^{\prime\mu\nu}$,
when $y\neq 0$
\begin{equation*}
M^3_5(\mathrm{P}^2-\partial_y^2)\tilde{h}_{\mu\nu}\tilde{T}^{\prime\mu\nu}=0
\end{equation*}

\begin{equation}
\tilde{h}_{\mu\nu}\tilde{T}^{\prime\mu\nu}=A e^{-\mathrm{P}\left|y\right|}
\end{equation}
The constant $A$ has to be determined now.
We can take the limit from both sides of the brane
\begin{equation*}
\lim_{\epsilon\to 0} \int\limits_{-\epsilon}^\epsilon{[M^3_5(\mathrm{P}^2-\partial_y^2)+M^2_4\delta(y)\mathrm{P}^2]\tilde{h}_{\mu\nu}\tilde{T}^{\prime\mu\nu}}\,dy=\int\limits_{-\epsilon}^\epsilon{(\tilde{T}_{\mu\nu}\tilde{T}^{\prime\mu\nu}-\frac{1}{3}\tilde{T}^\alpha_\alpha\tilde{T}^{\prime\beta}_\beta)\delta(y)}\,dy
\end{equation*}

\begin{equation*}
2M^3_5\mathrm{P}A+M_4^2\mathrm{P}^2A=\tilde{T}_{\mu\nu}\tilde{T}^{\prime\mu\nu}-\frac{1}{3}\tilde{T}^\alpha_\alpha\tilde{T}^{\prime\beta}_\beta
\end{equation*}

\begin{equation}
A=\frac{\tilde{T}_{\mu\nu}\tilde{T}^{\prime\mu\nu}-\frac{1}{3}\tilde{T}^\alpha_\alpha\tilde{T}^{\prime\beta}_\beta}{2M^3_5\mathrm{P}+M_4^2\mathrm{P}^2}
\end{equation}

Exactly on the brane, at $y=0$,

\begin{equation}
\label{note42}
\tilde{h}_{\mu\nu}(\mathrm{P},y=0)\tilde{T}^{\prime\mu\nu}(\mathrm{P})=\frac{\tilde{T}_{\mu\nu}\tilde{T}^{\prime\mu\nu}-\frac{1}{3}\tilde{T}^\alpha_\alpha\tilde{T}^{\prime\beta}_\beta}{2M^3_5\mathrm{P}+M_4^2\mathrm{P}^2}
\end{equation}

The potential mediated by this graviton on the brane is given as
\begin{equation}
V(r)=\int{h_{\mu\nu}T^{\mu\nu}(t,\mathbf{x},y=0;0,0,0)\,dt}
\end{equation}
where $h_{\mu\nu}T^{\mu\nu}$ is expressed in coordinate space.

Using \eqref{note42}, we find
\begin{equation}
V(r)=-\frac{1}{8\pi^2M^2_4r}\{\sin(\frac{r}{r_0})Ci(\frac{r}{r_0})+\frac{1}{2}\cos(\frac{r}{r_0})[\pi-2Si(\frac{r}{r_0})]\}
\end{equation}
where $Ci(z)\equiv\gamma+\ln(z)+\int_0^z(\cos(t)-1)\,dt/t$ is the Cosine Integral, $Si(z)\equiv\int_0^z\sin(t)\,dt/t$ is the Sine Integral, $\gamma$ is the Euler-Masceroni constant. $r_0$ is the characteristic distance scale defined as
\begin{equation}
r_0\equiv\frac{M_4^2}{2M_5^3}
\end{equation}
Two important limits are of course the short and long distance behavior of the potential.
When $r\ll r_0$, the potential is
\begin{align}
V(r) & \simeq-\frac{1}{8\pi^2M_4^2r}\{\frac{\pi}{2}+[-1+\gamma+\ln(\frac{r}{r_0})](\frac{r}{r_0}+\mathcal{O}(r^2))\}\\
& \sim -\frac{1}{r}
\end{align}
which gives the usual $4D$ Newtonian potential.
In the opposite limit, when $r\gg r_0$, the potential is
\begin{align}
V(r) & \simeq -\frac{1}{8\pi^2M_4^2}\frac{1}{r}[\frac{r_0}{r}+\mathcal{O}(\frac{1}{r^2})]\\
& \sim -\frac{1}{r^2}
\end{align}
The potential now scales as $1/r^2$, which is the regular $5D$ behavior. Thus, the DGP setup yields gravity that looks $4D$ on short and $5D$ on large distances.

\section{Extending DGP to lower dimensions}
\label{lddgp}

In this section, we will extend the DGP model to lower dimensions. In particular, our setup consists of a $3D$ brane embedded in a $4D$ space. We will rely on procedure reviewed in the previous section to calculate the behavior of the scalar and gravitational field in this framework.

\subsection{Scalar field}

We first concentrate on the scalar field propagator. The scalar field Lagrangian that we want to study is
\begin{equation}
\mathcal{L}=-\frac{M_4^2}{2}\partial_\mu\phi\partial^\mu\phi-\frac{M_3}{2}\delta(z)\partial_i\phi\partial^i\phi
\end{equation}
where lowercase Latin indices $i$, $j$, $k$,$\cdots$ are used for $3D$ quantities, while $z$ is reserved for the bulk coordinate.
The Green's function for this case is
\begin{equation}
(M_4^2\partial_\mu\partial^\mu+M_3\delta(z)\partial_i\partial^i)G(x,z;0,0)=\delta^3(x)\delta(z)
\end{equation}
If we perform the Fourier transformation and work in Euclidean space, we get
\begin{equation}
[M_4^2(p^2-\partial^2_z)+M_3p^2\delta(z)]\tilde{G}(p,z)=\delta(z) .
\end{equation}
Solving this equation we find
\begin{equation}
\tilde{G}(p,z)=\frac{1}{2M_4^2p+M_3p^2} e^{-p \left|z\right|} .
\end{equation}
Thus the potential mediated by this scalar field on the $3D$ brane is,
\begin{equation}
V(r)=-\frac{1}{8\pi^2M_4^2}\frac{1}{r}\{\pi-2Ci(\frac{r}{r_0})\sin(\frac{r}{r_0})-\cos(\frac{r}{r_0})[\pi-2Si(\frac{r}{r_0})]\}
\end{equation}
where $r_0=\frac{M_3}{M_4^2}$ is the critical distance.
At short distances, where $r\ll r_0$, the potential is
\begin{align}
V(r) & \simeq -\frac{1}{8\pi^2M_4^2}\frac{1}{r}\{-2[-1+\gamma+\ln(\frac{r}{r_0})]r+O(r^2)\} \\
& \sim -\ln(\frac{r}{r_0})
\end{align}
The $\ln(r)$ behavior is what we expect in a $3D$  space.
At large distances, where $r\gg r_0$, the potential is
\begin{align}
V(r) & \simeq -\frac{1}{8\pi^2M_4^2}\frac{1}{r}[\pi-2\frac{r_0}{r}+O(\frac{1}{r^2})] \\
& \sim -\frac{1}{r}
\end{align}
Obviously, the standard $4D$ behavior is recovered on large distances.

\subsection{Gravity}
\label{gravity}

To learn about the tensor structure of this setup, we have to write down gravitational action.
We define the stress-energy tensor on the $3D$ brane as
\begin{equation}
T_{\mu\nu}= \left(\begin{array}{cc}
\delta(z)T_{ij}&0\\
0&0
\end{array} \right)
\end{equation}
while the action is,
\begin{equation}
S=\frac{M_4^2}{2}\int{\mathrm{d}^4x\,\sqrt{-g}R_4}+\frac{M_3}{2}\int{\mathrm{d}^3x\,\sqrt{-g_3}R_3}
\end{equation}
The equations of motion take the form
\begin{equation} \label{eom}
\frac{M_4^2}{2}G_{\mu\nu}+\frac{\delta(z)}{2}M_3G_{ij}\delta^i_\mu\delta^j_\nu=\delta(z)T_{ij}\delta^i_\mu\delta^j_\nu
\end{equation}
With the choice of harmonic gauge in the $4D$ bulk, one can check that $h_{i4}=0$ and  $\partial^i h_{ij}=\partial_j h^k_k$ are consistent with the equations of motion.
In turn, this leads to the $(ij)$ components of the Eq.~(\ref{eom})
\begin{equation}
-\frac{1}{2}(M_4^2\square + \delta(z)M_3\square_3)h_{ij}=(T_{ij}-\frac{1}{2}\eta_{ij}T)\delta(z)-\frac{M_3}{2}\delta(z)\partial_i\partial_j h^4_4
\end{equation}
Performing the Fourier transform in Euclidean space as before we get
\begin{equation}
\tilde{h}_{ij}\tilde{T}^{\prime ij}=\frac{\tilde{T}_{ij}\tilde{T}^{\prime ij}-\frac{1}{2}\tilde{T}_k^k\tilde{T}_l^{\prime l}}{2M_4^2p +M_3p^2}
\end{equation}
From this equation we can read the tensor structure of the lower dimensional DGP. Obviously, it is different from Eq.~(\ref{note42}), but the difference comes only from the reduced dimensionality. The importance of this result is that DGP can be extended to lower dimensions without any fundamental problems.

\section{Cascading scalar model $5D$ $\rightarrow$ $4D$ $\rightarrow$ $3D$}
\label{cascadingscalar}
All that we have achieved so far is to merely extend the DGP model to lower dimensions, and unfortunately will carry on a bad character from the DGP framework - unstable high energy modes.  One way that can take us out of this crisis is to successively embed the codimension-2 brane into a codimension-1 brane and then to embed both of them in a bulk, in analogy with  \cite{d,g,h}. In this section, we will examine this idea in our lower dimensional scenario, with a scalar field model. In the next section we will do the same with gravity.

Following the procedure in \cite{h}, the full action in our context is

\begin{equation} \label{scalaraction}
S= -\frac{M_5^3}{2}\int{\mathrm{d}^5x\,\partial_A\phi\partial^A\phi}-\frac{M_4^2}{2}\int{\mathrm{d}^4x\,\partial_\mu\phi\partial^\mu\phi}-\frac{M_3}{2}\int{\mathrm{d}^3x\,\partial_i\phi\partial^i\phi}
\end{equation}

To find the final propagator on the codimension-2 brane, we treat the first term in \eqref{scalaraction} as the free theory and the rest as interactions with coupling $\lambda_1=-M_4^2(p^2+q_1^2)$ on the codimension-1 brane and coupling $\lambda_2=-M_3P^2$ on the codimension-2 brane. Here $q_1$ is the momentum in the $z$ dimension and $p_i$ the momentum on the codimension-2 brane.

In momentum space, the free propagator on the codimension-1 brane is,
\begin{equation}
G_0(p_i,q_1;0,0)=\frac{1}{M_5^3}\int{\frac{dq}{\pi}\frac{1}{p^2+q_1^2+q^2}}=\frac{1}{M_5^3\sqrt{p^2+q_1^2}} .
\end{equation}

This propagator couples with the first interaction term and produces the modified propagator on the codimension-1 brane
\begin{equation}
\begin{split}
G_1(p_i,q_1) &= G_0(p_i,q_1)+\lambda_1G_0(p_i,q_1)^2+\lambda_1^2G_0(p_i,q_1)^3+ \cdots \\
&= \frac{G_0(p_i,q_1)}{1+M_4^2(p^2+q_1^2)G_0(p_i,q_1)}\\
&= \frac{1}{M_4^2}\frac{1}{p^2+q_1^2+m_5\sqrt{p^2+q_1^2}}
\end{split}
\end{equation}
where $m_5$ is the $5D$ to $4D$ crossover scale \[ m_5=\frac{M_5^3}{M_4^2}\].

On the codimension-2 brane we have,
\begin{equation}
\begin{split}
G_1(p_i)\equiv G_1(p_i;0,0) &= \frac{1}{M_4^2}\int{\frac{dq_1}{\pi}\frac{1}{p^2+q_1^2+m_5\sqrt{p^2+q_1^2}}}\\
&= \frac{4}{\pi M_4^2}\frac{1}{\sqrt{p^2-m_5^2}}\tan^{-1}(\sqrt{\frac{p-m_5}{p+m_5}})
\end{split}
\end{equation}

Note that when $p<m_5$, $\tan^{-1}(\sqrt{\frac{p-m_5}{p+m_5}})=i\tanh^{-1}(\sqrt{\frac{m_5-p}{p+m_5}})$.

If we carry on the same procedure as shown above with the second coupling constant $\lambda_2=-M_3P^2$, we arrive at the final propagator on the codimension-2 brane:
\begin{equation}
G_2(p_i) =\frac{G_1(p_i)}{1+M_3p^2G_1(p_i)}=\frac{1}{M_3}\frac{1}{p^2+g(p^2)}
\end{equation}
where $g(p^2)$ is defined for future reference as,
\begin{equation}
\label{functiong}
g(p^2)\equiv  \frac{\pi M_4^2}{4}\frac{\sqrt{p^2-m_5^2}}{\tan^{-1}(\sqrt{\frac{p-m_5}{p+m_5}})} \qquad  for \quad p>m_5
\end{equation}
with $m_4=\frac{M_4^2}{M_3}$ the $4D$ to $3D$ crossover scale.

It is worth noting that all what is required for a well defined propagator here is to have a non-zero Planck mass $M_4$. The existence of the codimension-1 brane serves as a regulator, without which the propagator is divergent
\begin{equation}
\mbox{if $M_4 \rightarrow 0$ or $m_5\gg p$},\quad G_2(p_i)\longrightarrow \frac{1}{2M_5^3}\log(\frac{p}{m_5})
\end{equation}
On the other hand, if the scalar field does not feel the $5th$ dimension, i.e. $m_5\rightarrow0$, then the DGP propagator is recovered,
\begin{equation}
G_2(p_i)\longrightarrow \frac{1}{M_3}\frac{1}{p^2+m_4p}
\end{equation}
Therefore a successive embedding of codimension 1 and 2 branes is essential in the set-up of cascading models.

\section{Cascading gravity $5D$ $\rightarrow$ $4D$ $\rightarrow$ $3D$}

Generalization to the gravitational sector is also straightforward, but unlike the scalar sector, in the gravitational sector a scalar mode appears to be the source of instability. The same features are found in the original \emph{cascading models} \cite{d,g,h}, and the same cure applies here as well.

Starting from the $5D$ action,
\begin{equation}
S= \frac{M_5^3}{2}\int{\mathrm{d}^5x\,\sqrt{-g_5}R_5}+\frac{M_4^2}{2}\int{\mathrm{d}^4x\,\sqrt{-g_4}R_4}+\frac{M_3}{2}\int{\mathrm{d}^3x\,\sqrt{-g_3}R_3} ,
\end{equation}
the $5D$ Einstein equations are what is previously seen in Eq. \eqref{23inDGP}, section \ref{subsection2b} :
\begin{equation}
-\frac{M_5^3}{2}(\square_5+\frac{\delta(y)}{m_5}\square_4)h_{\mu\nu}=(T_{\mu\nu}^{(4)}-\frac{1}{3}\eta_{\mu\nu}T^{(4)})\delta(y) -\frac{\delta(y)}{2}M^2_4\partial_\mu\partial_\nu h^\gamma_\gamma
\end{equation}
With the presence of the codimension-2 brane, the $4D$ stress-energy tensor takes the form,
\begin{equation}
T_{\mu\nu}^{(4)}=\delta(z)(-M_3G_{ij}^{(3)}+T_{ij}^{(3)})\delta_\mu^i\delta_{\nu}^j
\end{equation}

Here and henceforth we use the superscript number in parenthesis to denote the quantities in corresponding dimensions.

Suppose the space-time  on the $3D$ brane is flat, then the $3D$ perturbations can be decomposed into a scalar($\pi$) part and a transverse and traceless(TT) part,
\begin{equation}
\label{decomposition}
h_{ij}^{(3)}=h_{ij}^{(3)TT}+\pi\eta_{ij}
\end{equation}
The $4D$ transverse and traceless variable can also be projected out as
\begin{equation}
\label{4dTT}
h_{\mu\nu}^{(4)TT}=h_{\mu\nu}-\frac{\partial_\mu\partial_\nu}{\square_4}h^\gamma_\gamma
\end{equation}
which leads to a decoupled and regularized equation,
\begin{equation}
-\frac{M_5^3}{2}(\square_5+\frac{\delta(y)}{m_5}\square_4)h_{\mu\nu}^{(4)TT}=\delta(y)(T_{\mu\nu}^{(4)}-\frac{1}{3}T^{(4)}\eta_{\mu\nu}+\frac{1}{3}\frac{\partial_\mu\partial_\nu}{\square_4}T^{(4)})
\end{equation}
This equation can be solved and the result takes the form,
\begin{equation}
\label{4Deq}
-\frac{M_4^2}{2}(\square_4-m_5\sqrt{-\square_4})h_{\mu\nu}^{(4)TT}=\delta(z)(T_{ij}^{(3)}-M_3G_{ij}^{(3)})\delta_\mu^i\delta_\nu^j-\frac{\delta(z)}{3}(\eta_{\mu\nu}-\frac{\partial_\mu\partial_\nu}{\square_4})(T^{(3)}+\frac{1}{2}M_3R_3)
\end{equation}
where $h_{\mu\nu}^{(4)TT}$ is recognized as  the propagator on the codimension-1 brane.

Just as in the scalar case, the perturbation on the codimension-2 brane can be obtained when the interaction there is taken into account. With our choice $h_{ij}^{(3)}=h_{ij}^{(3)TT}+\pi\eta_{ij}$, it can be shown that
\begin{equation}
R_{ij}^{(3)}=-\frac{1}{2}(\partial_i\partial_j\pi+\square_3h_{ij}^{(3)TT}+\square_3\pi\eta_{ij})
\end{equation}

\begin{equation}
R_3=-2\square_3\pi
\end{equation}

\begin{equation}
G_{ij}^{(3)}=-\frac{1}{2}[\square_3h_{ij}^{(3)TT}-\square_3(\eta_{ij}-\frac{\partial_i\partial_j}{\square_3})\pi]
\end{equation}

One can fix the gauge in which  
\begin{equation}
\label{pi}
\pi=\frac{\square_4}{2\square_3}\eta^{ij}h_{ij}^{(4)TT}=-\frac{\square_4}{2\square_3}h_{zz}^{(4)TT} ,
\end{equation}
with which $h_{ij}^{(3)TT}$ can be expressed in terms of $h_{ij}^{(4)TT}$ and $\pi$ as
\begin{equation}
h_{ij}^{(3)TT}=h_{ij}^{(4)TT}-\pi\eta_{ij}-(\frac{2\square_3}{\square_4}-3)\frac{1}{\square_3}\partial_i\partial_j\pi .
\end{equation}
The $(ij)$ component of \eqref{4Deq} gives us an equation for the TT part
\begin{equation}
-\frac{M_4^2}{2}(\square_4-m_5\sqrt{-\square_4})h_{ij}^{(3)TT}=\delta(z)(\Sigma_{ij}^{(3)}+\frac{M_3}{2}\square_3h_{ij}^{(3)TT})
\end{equation}
where $ \Sigma_{ij}^{(3)}=T_{ij}^{(3)}-\frac{1}{2}\eta_{ij}T^{(3)}+\frac{1}{2}\frac{\partial_i\partial_j}{\square_3}T^{(3)} $ is the TT part of the matter-stress tensor.

Comparing this equation with the equation of motion obtained in the Section \ref{cascadingscalar} for the scalar model, it is clear that they take the same general form. The solution of $h_{ij}^{(3)TT}$ is,
\begin{equation}
h_{ij}^{(3)TT}=\frac{2}{M_3}\frac{1}{-\square_3+g(-\square_3)}\Sigma_{ij}^{(3)}
\end{equation}
where function $g$ was defined in Eq. \eqref{functiong}. The $(zz)$ component of Eq.~(\eqref{4Deq}) yields an equation for $\pi$

\begin{equation}
-\frac{M_4^2}{2}(\square_4-m_5\sqrt{-\square_4})\pi=\frac{1}{6}\delta(z)(T^{(3)}-M_3\square_3\pi)
\end{equation}
from which the solution is found to be,
\begin{equation}
\pi=\frac{1}{M_3}\frac{1}{\square_3+3g(-\square_3)}T^{(3)} .
\end{equation}

This solution is problematic because, unlike the scalar solution and the transverse traceless tensor modes, the overall sign in front of $\square_3$ appears to be positive, and thus at high energy this propagator is not positive definite. However such a ghost is generic in
any codimension-2 theory, and within the framework of  \emph{cascading gravity}, two cures have been proposed, both aimed at providing an additional positive kinetic term for $\pi$. The first approach is to introduce a high enough tension to the codimension-2 \cite{d}, which produces extrinsic curvature while keeping the metric on the brane flat. The net result is the $\pi$ field obtains contributions from the additional terms in the lagrangian involving tension  and a healthy sign is restored for $\pi$. An alternative approach is to regularize the codimension-2 brane with the understanding that an $R_3$ is not the most general induced gravity action \cite{h}. Different regularization methods at the end yield the same tensor scalar result with an additional kinetic term for $\pi$.

Using the general formula derived in \cite{h}, the trace of the effective source can be found,
\begin{equation}
T_{eff}=\delta(z)(-\frac{3}{2}M_3R_3+T^{(3)}) .
\end{equation}
In our decomposition \eqref{decomposition}, $R_3=-2\square_3\pi$, thus the regularized equation of motion for $\pi$ takes the form,
\begin{equation}
-\frac{M_4^2}{2}(\square_4-m_5\sqrt{-\square_4})\pi-\frac{M_3}{2}\delta(z)\square_3\pi=\frac{1}{6}\delta(z)T^{(3)}
\end{equation}
with a healthy solution,
\begin{equation}
\pi=\frac{1}{3M_3}\frac{1}{-\square_3+g(-\square_3)}T^{(3)}
\end{equation}

This result indicates that the cascading gravity model can be self-consistently extended to lower dimensions.

It is interesting to note that we obtained non-trivial results when extending the cascading gravity models to lower dimensions, though it is well known that the $2+1$ dimensional space-times in the general relativity do not have local propagating degrees of freedom (gravity waves in classical theory or gravitons in quantum theory). However, the tensorial structure of the DGP models is such that graviton on the brane acquires an induced mass, so they it is not massless like in the general relativity. And massive gravity is known to have propagating degrees of freedom even in $2+1$ dimensional space-times. This magic would however stop in $1+1$ dimensional space-time, where Einstein's action is just an Euler's characteristic of the manifold in question.

\section{Possible applications}

\begin{figure}[h!]
   \centering
\includegraphics[width=8cm]{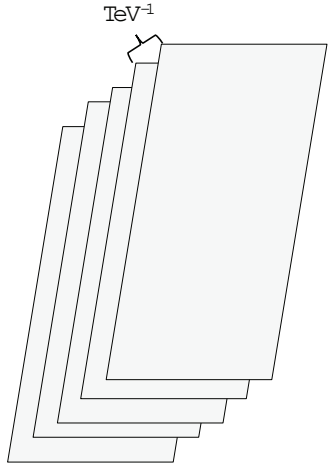}
\caption{The 4D brane that we live on could be comprised of the stack of densely packed 3D branes. If the inter-brane separation corresponds to the dimensional cross-over scale (TeV$^{-1}$ or smaller), our 4D universe would appear smooth at energies smaller than TeV.}
\label{stack}
\end{figure}

One of the possible applications of our setup is the framework of ``vanishing dimensions". In this framework, high energy physics is lower dimensional while the low energy physics is higher dimensional. Dimensions open up one by one as we increase the length scale. On the usual length scales  between $10^{-17}$cm and a Gpc, our universe is 4D, as we well know. On scales comparable to out cosmological horizon and lager, our universe may become 5D and so on. On scales shorter than $10^{-17}$cm, our universe may appear 3D. In order to achieve this, we have to slightly modify our setup. In our original setup we have only one 3D brane embedded in a 4D brane (which is itself embedded in a 5D brane and so on). This would make our universe appear lower dimensional only at one particular point, which is the location of the 3D brane. In order to make our 4D universe appear lower dimensional around every point, we have to introduce a stack of branes with the inter-brane distance that would correspond to the dimensional cross-over scale (TeV$^{-1}$ or $10^{-17}$cm is the maximal value is required by phenomenology). Then the 4D brane that we live on, would actually be comprised of the many densely packed 3D branes, as in Fig.~\ref{stack} (this work will appear in \cite{Dai:2014roa}). At energies smaller than the inverse dimensional cross-over scale our 4D space would appear smooth.

This modified setup would require adding a set of N lower dimensional branes instead of just one. In particular, the term $\delta (z)$ in Section \ref{gravity} and the subsequent sections would need to be replaced with
\be
\sum_i^N \delta (z-z_i) ,
\ee
where $z_i$ are the locations of the 3D branes inside the 4D  one. This setup would resemble an anisotropic lattice model in \cite{Anchordoqui:2010er}. At energies higher than the dimensional crossover scale, one may encounter interesting effects like planar scattering, eliptic jets etc.

\section{Conclusions}

In this paper we showed that the cascading gravity model can be successfully extended to lower dimensions. In particular, one can self-consistently embed a 3D brane into a 4D brane which is itself embedded into a 5D brane. We demonstrated this on the example of the scalar field and graviton. The same mechanism which cures instability in the case of gravity in a higher dimensional cascading gravity model works in our case too.

Apart from the pure academic interest,  our setup fits well into the ``vanishing dimensions" framework where the short distance physics appears to be lower dimensional.

\acknowledgments

This work was also partially supported by the US National Science Foundation, under Grant No. PHY-1066278 and PHY-1417317.
DS also acknowledges hospitality and support from the Perimeter Institute for Theoretical Physics.


\end{document}